\documentstyle[12pt]{article}
\newcommand{\be}{\begin{equation}}
\newcommand{\ee}{\end{equation}}
\newcommand{\bea}{\begin{eqnarray}}
\newcommand{\eea}{\end{eqnarray}}
\newcommand{\bdm}{\begin{displaymath}}
\newcommand{\edm}{\end{displaymath}}
\title{\Large Asymptotic States in Non-Local Field Theories}
\author{D.\ G.\ Barci\thanks{barci@symbcomp.uerj.br} \\ Universidade do
Estado do Rio de Janeiro,\\
Instituto de F\'\i sica, Departamento de F\'\i sica Te\'orica \\
R.\ S\~ao Francisco Xavier, 524,\\ Maracan\~a, cep 20550, Rio de Janeiro,
Brazil.
\and
L.\ E.\ Oxman\thanks{oxman@if.ufrj.br} \\Instituto de F\'\i sica,
 Departamento de F\'\i sica Te\'orica\\
Universidade Federal do  Rio de Janeiro \\
C.P. 68528, Rio de Janeiro, RJ, 21945-970, Brazil. }
\date{July 7, 1996}
\begin{document}
\maketitle
\begin{abstract}
Asymptotic states in field theories containing non-local kinetic terms are
analyzed using the canonical method, naturally defined in Minkowski space.

We apply our results to study the asymptotic states of a non-local
Maxwell-Chern-Simons theory coming from bosonization in $2+1$
dimensions. We show that in this case the only asymptotic state of
the theory, in the trivial (non-topological) sector, is the vacuum.
\end{abstract}
\newpage

Field theories containing non-local kinetic terms are receiving increasing
attention, in relation with different motivations.

This kind of non-locality appears, for example, when considering effective
field theories where some degrees of freedom have been integrated out in
an underlying local field theory.

In Ref.\ \cite{marino-proyeccion}, it was shown that the kinetic term of QED,
obtained by projection from a $(3+1)-$dimensional space to a
$(2+1)-$dimensional one, is proportional to $F^{\mu\nu}\Box^{-1/2}F_{\mu\nu}$.
In the static
limit, this term reproduces correctly the $1/r$ Coulomb potential
instead of the usual logarithmic behavior of $(2+1)$D QED, this fact
was first noticed in Ref. \cite{dalam1/2}.

Non-local kinetic terms also appear in the context of bosonization of massless
fermions in $(2+1)$ dimensions.
In Ref.\ \cite{marino}, the following mapping has been established
\begin{eqnarray}
\bar{\psi}\not \! \partial \psi &\leftrightarrow&
\frac{1}{4}\, F_{\mu\nu}(-\partial^2)^{-1/2}F_{\mu\nu}
\,+\,\frac{i}{2}\,\theta \,\epsilon_{\mu\nu\lambda}A_\mu\partial_\nu
A_\lambda \,+\, nqt \nonumber \\
\bar{\psi}\gamma_\mu\psi &\leftrightarrow&
\beta \, \epsilon_{\mu\nu\lambda}\partial_\nu A_\lambda
\,-\,\beta \,\theta \,(-\partial^2)^{-1/2} \partial_\nu F_{\mu\nu}
{}~~,
\label{1.1}
\end{eqnarray}
where $\psi$ is a two-component Dirac spinor, $A_\mu$ is a $U(1)$
gauge field, and $nqt$ are non-quadratic terms.  The parameter
$\theta$ is regularization-dependent.

More recently, in Ref.\ \cite{bfo}, we have extended the path-integral
bosonization scheme proposed in Ref.\ \cite{fidel}, for massive fermions in
$2+1$ dimensions, obtaining a non-local bosonized gauge theory.
This enabled us to discuss both the massive and massless fermion cases on an
equal footing.

By using this kind of non-local extension, the Schwinger terms associated with
fermions in $2+1$ dimensions have been computed in Ref.\ \cite{fln}; also, in
Ref.\ \cite{bm}, the bosonization of the massive Thirring model in four
dimensions was considered.

These relations are studied, in general, by following path integral techniques,
in a way similar to the local case.

The development of canonical methods, suited to study the type of non-locality
that appears (for example) in the context of bosonization, is recent.

In Ref.\ \cite{amaral-marino}, the Dirac quantization of non-local theories
containing fractional powers of the d'alambertian operator was considered.
In that work, the interpretation of non-local theories as infinite order
derivative theories leads to an infinite set of canonical momenta which satisfy
an infinite set of second class constraints.

In Ref.\ \cite{bor}, we have carried out an alternative approach to
(canonically) quantize theories containing general non-local kinetic terms, by
using the Schwinger method. This technique has the advantage of avoiding the
definition of (infinite) canonical momenta. It is essentially based on the
observation that the hamiltonian, conserved due to time translation symmetry,
must be the generator of time translations. In this way, the canonical
commutation relations can be deduced from Heisenberg's equation.
In order to implement this approach, we obtained, in Ref.\ \cite{bor}, a mode
expansion for the solutions to the classical (non-local) homogeneous equations
as well as a closed expression for the conserved hamiltonian. In that
reference, we have also shown examples of theories containing non-local kinetic
terms where the hamiltonian is positive definite. This is in contrast with the
higher order case, where the hamiltonian is not positive definite, leading, in
general, to instability problems and the related unitarity problem.

In Ref. \cite{psdif}, we have established a consistent mathematical formalism
to deal with non-local kinetic operators, where the solutions to the non-local
equations are shown to belong to a space of ultradistributions.
These solutions display a continuum of massive modes.

The aim of this letter is to study the asymptotic states in the trivial sector
of field theories containing non-local kinetic terms; this will contribute to
understand the physical content of these theories. We will also apply our
technique to the gauge theory obtained in the context of $(2+1)$D bosonization
(see Ref. \cite{bfo}), showing that the proposed non-local bosonized lagrangian
defines a sensible theory in Minkowski space.

Before analyzing the asymptotic states of a non-local theory, we will
review the above mentioned quantization procedure (see  Ref.\ \cite{bor}).

Let us consider the non-local lagrangian for a real scalar
field $\phi(x)$
\begin{equation}
{\cal L}=\frac{1}{2}\phi f(\Box) \phi~~~,
\label{L}
\end{equation}
where $f(z)$ is a general analytic function having a
cut contained in the negative real axis.
The corresponding equation of motion is:
\begin{equation}
f(\Box) \phi = 0~~~.
\label{EM}
\end{equation}
and a general solution can be written in the following form \cite{bor}
\cite{psdif}
\begin{equation}
\phi (x)=i\int_{\Gamma_+ +\Gamma_-}dk e^{ikx}\frac{1}{f(-k^2)} a(k)
=\int dk [e^{ikx}a(k)+e^{-ikx}\bar{a}(k)]\delta^+ G (k^2)~~~.
\label{fic}
\end{equation}
In the first equality, $dk$ integrates over the whole space-time,
with $k_0$ moving along $\Gamma_+ +\Gamma_-$ where $\Gamma_+$
(resp. $\Gamma_-$) is a path surrounding in the positive (resp. negative)
sense, all the singularities of $1/f(-k^2)$, which are present in the
positive (resp. negative) $k_0$-axis. $a(k)=a(k_0,{\bf k})$ is an entire
analytic function in the $k_0$ variable; when we apply $f(\Box)$ on eq.\
(\ref{fic}), the integrand becomes an entire analytic function in $k_0$ and the
integral is zero, showing that (\ref{fic}) is a solution to eq.\ (\ref{EM}).

In the second equality we have written the solution in terms of the real
distribution (weight function) $\delta^+ G (k^2)$ given by
\begin{equation}
\delta^+ G (k^2)\equiv -i\, \left[
\frac{1}{f(-(k_0+i\epsilon)^2+{\bf k}^2)}-
\frac{1}{f(-(k_0-i\epsilon)^2+{\bf k}^2)}\right] \theta (k_0)~~~.
\label{d+}
\end{equation}

 From Noether's theorem, we can deduce (see Ref.\ \cite{bor})
\begin{equation}
H=\int dk\, k_0\, \delta^+ G (k^2)
a^{\dagger}(k_0,{\bf k})a(k_0,{\bf k})~~~.
\label{Hrq}
\end{equation}
Now, at the quantum level, the field operator must satisfy Heisenberg's
equation. This implies the algebra ($k_0>0$, $k'_0>0$):
\begin{equation}
\delta^+ G(k^2)\delta^+ G(k'^2) [a(k),a^{\dagger}(k')]=\delta^+ G(k^2) \delta
(k-k')
\makebox[.5in]{,}
[a(k),a(k')]=0~~~.
\label{algebra}
\end{equation}
The field commutator is
\begin{equation}
\Delta(x-y)=[\phi(x),\phi(y)]=i \int_{\Gamma} dk \frac{e^{ikx}}{f(-k^2)}~~~.
\end{equation}
In terms of the weight function,  this equation reads
\begin{equation}
\Delta(x-y)=\int d\mu~ \delta^+G(\mu) \Delta_{\mu^2}(x-y)~~~,
\label{commutator}
\end{equation}
where $\Delta_{\mu^2}(x-y)$ is the Pauli-Jordan distribution with mass
parameter $\mu$.

Now, from the commutator (\ref{commutator}), we see that the field associated
with the non-local lagrangian (\ref{L}) is  a {\em generalized free field},
 as introduced in Ref.\   \cite{green}.
In that work it is shown that the asymptotic states (in the vacuum
sector) come from the $\delta$-singularities of the weight
function $\delta^+G$.

Precisely, for a general Heisenberg field $\phi(x)$ be associated with
asymptotic
modes of mass $m$ is necessary that the  {\it in-out} fields
defined by
\begin{equation}
\phi_{m^2}^{ {\tiny\matrix{out\cr in}}}(x)=\lim_{\tau\rightarrow \pm\infty}
\int_{y^0=\tau} d{\bf y}~\left[ \phi(y)\frac{\stackrel{\leftrightarrow}
{\partial}~}{\partial y^0}\Delta_{m^2}(x-y)\right]
\label{inout}
\end{equation}
($\phi\frac{\stackrel{\leftrightarrow}{\partial}~}{\partial y^0}
\psi=\phi \frac{\partial\psi}{\partial y^0}-\psi \frac{\partial\phi}
{\partial y^0}$)
have free field commutation relations
\begin{equation}
[\phi_{m^2}^{{\tiny\matrix{out\cr in}}}(x),\phi_{m^2}^{{\tiny
\matrix{out\cr in}}}(y)]
=i\Delta_{m^2}(x-y)~~~.
\label{com}
\end{equation}

If $\phi$ is a free field with mass $m_0$, it is easy to show,
using the properties
\begin{equation}
\lim_{\tau\rightarrow \pm\infty}
\int_{y^0=\tau} d{\bf y}~ \Delta_{m_0^2}(x-y)\frac{\stackrel{\leftrightarrow}
{\partial}~}{\partial y^0}\Delta_{m^2}(z-y)=
\left\{\matrix{\Delta_{m_0^2}(x-z),~~\mbox{if~~}m^2=m_0^2\cr
			0  ~~~~~~~~~~~~~~\mbox{otherwise} } \right.~~~,
\label{property}
\end{equation}
that  the commutator (\ref{com}) is the free commutator
$\Delta_{m^2_0}(x-y)$, if $m^2=m_0^2$, while it is zero when $m^2\neq m_0^2$,
as expected.

In the case of a non-local field, the asymptotic states only come from the
$\delta$-singularities of the weight function $\delta^+G$ (poles of $1/f$).
This comes about from eq.\ (\ref{fic}), which expresses the on-shell field as a
superposition of free fields with mass $k^2$. If we consider in eq.\
(\ref{inout}) an $m^2$ value where $\delta^+G$ is a well defined function,
the commutator (\ref{com}) will be zero
(the mass mode $m^2$ has zero measure in eq.\ (\ref{inout})). In particular, if
$1/f$ only has cuts, then, the unique asymptotic state (in the trivial sector)
is the vacuum.

Similarly, at the classical level, the field modes in the continuum (where
$\delta^+G$ is a well defined function) cannot be
associated with asymptotic free wave solutions representing a particle with
mass $m^2$, that is, there is no solution such as $\exp{ikx}$, $k^2=m^2$;
remember that $a^{\mu}(k_0,\vec{k})$ in (\ref{fic}) cannot be a Dirac delta, as
it must be an entire analytic function of $k_0$.

Let us now show how this result clarifies  the analysis of asymptotic states
of a non-local gauge theory.

In reference \cite{bfo}, we extended a recently proposed  path-integral
bosonization scheme for massive fermions in $2+1$ dimensions \cite{fidel}
by keeping the full momentum-dependence of the one-loop vacuum polarization
tensor. This enabled us to discuss both the massive and massless
fermion cases on an equal footing.

In that reference we have shown that the generating functional that bosonizes
the fermionic currents of free massive fermions in $2+1$ dimensions is
\bea
Z(s)&=&\int [dA]
\exp -\int d^3 x
[ \frac{1}{4} F_{\mu \nu} \, C_1 \, F_{\mu \nu}
- \frac{i}{2} A_{\mu} \, C_2 \, \epsilon_{\mu \nu \lambda}
\partial_{\nu} A_{\lambda}  \nonumber\\
&+& i \, (\frac{u_+ - u_-}{2}) \, s_\mu
\frac{1}{\sqrt{-\partial^2}} \partial_{\nu} F_{\nu \mu}
\,-\, i \, (\frac{u_+ + u_-}{2})\, s_\mu \epsilon_{\mu \nu \lambda}
\partial_{\nu} A_{\lambda} ]~~~,
\label{1.19}
\eea
where
\bea
C_1 &=& \frac{1}{2} \; \frac{ |u_+|^2 (F - i G) \,+\,|u_-|^2
(F + i G)}{ - \partial^2 F^2 \, + \, G^2 } \nonumber\\
C_2 &=& \frac{i}{2} \; \frac{ |u_+|^2 (F - i G) \,-\,|u_-|^2
(F + i G)}{ - \partial^2 F^2 \, + \, G^2 }~~~,
\label{1.20}
\eea
$F$ and $G$ come from the parity conserving and parity violating
parts of the fermionic one-loop vacuum polarization tensor, respectively;
$u_\pm$ are arbitrary parameters associated with a redefinition of the
(dummy) gauge field $A_\mu$ in eq.\ (\ref{1.19}).
In (Euclidean) momentum representation they are given by
\be
{\tilde F} \;=\; \frac{\mid m \mid}{4 \pi k_E^2} \,
\left[ 1 - \displaystyle{\frac{1 \,-\,\displaystyle{\frac{k_E^2}{4 m^2}}}{(
\displaystyle{\frac{k_E^2}{4 m^2}})^{\frac{1}{2}}}} \, \arcsin(1\,+
\, \frac{4 m^2}{k_E^2})^{-\frac{1}{2}} \right] ~~~,
\label{1.10}
\ee
the function ${\tilde G}$ in (\ref{1.20}) is regularization dependent, and
can be written as
\be
{\tilde G} \;=\; \frac{q}{4 \pi} \,+\, \frac{m}{2 \pi \mid k_E \mid}
\, \arcsin (1 \, + \, \frac{4 m^2}{k_E^2} )^{- \frac{1}{2}} ~~~,
\label{1.11}
\ee
where $q$ can assume any integer value, and may be
thought of as the effective number of Pauli-Villars regulators, namely, the
number of regulators with positive mass minus the number of
negative mass ones.

When $m\rightarrow\infty$ (taking $u_+=u_-=\frac{1}{2\pi}$)
\begin{equation}
S_{bos} \,=\, \int d^3 x \, \left( \pm \frac{i}{2}
A_{\mu} \epsilon_{\mu \nu \lambda} \partial_{\nu} A_\lambda
\,-\, \frac{i}{\sqrt{4 \pi}} s_{\mu} \epsilon_{\mu \nu \lambda}
\partial_{\nu} A_{\lambda} \right) ~~~,
\label{1.22}
\end{equation}
which agrees with the result of \cite{fidel}.

When $m\rightarrow 0$ (taking $u_+=u_-^\ast=(\exp{i\alpha})/4$)
\bea
S_{bos} &=& \int d^3 x \, ( \frac{1}{4} \,
F_{\mu \nu} \frac{1}{\sqrt{-\partial^2}} F_{\mu \nu} \;-\;
\frac{i}{2} \, \frac{\pi}{4 q} \, \epsilon_{\mu \nu \lambda}
A_\mu \partial_\nu A_\lambda \nonumber\\
&-& \frac{ \sin \alpha }{4} \, s_\mu \, \frac{\partial_\nu
F_{\nu \mu}}{\sqrt{-\partial^2}} \;-\; i \frac{\cos \alpha}{4}
\, \epsilon_{\mu \nu \lambda} \partial_\nu A_\lambda ) ~~~,
\label{1.26}
\eea
This result coincides with that obtained in \cite{marino} (cf.\ eq.\
(\ref{1.1})).

The generating functional (\ref{1.19}) is a non-local extension of the local
Maxwell-Chern-Simons theory studied in Ref. \cite{templeton}.
Now, let us analyze the {\em asymptotic states} in the trivial
(non-topological) sector of the theory.

Going back to Minkowski space, and setting the currents equal to zero,
we obtain from (\ref{1.19}) the homogeneous Euler-Lagrange equation
associated with the bosonized action
\be
 C_1 \partial_\nu F_{\nu \mu}
+ C_2  \epsilon_{\mu \nu \lambda}
\partial_{\nu} A_{\lambda}=0~~~.
\label{ll1}
\ee
where $C_1$ and $C_2$ are the functions of the d'alambertian operator {\it
defined in Minkowski space} and obtained from eqs.\ (\ref{1.20}), (\ref{1.10})
and (\ref{1.11}) by the replacement $-\partial^2\rightarrow \Box=
\partial_0^2-\nabla^2~,~-k_E^2\rightarrow k^2=k_0^2-{\bf k}^2$.

This equation may be written, in the equivalent form
\be
C_1^{-1}\left(C_1^2\Box+C_2^2\right)~^* F^\mu=0~~~,
\label{dual}
\ee
where $^*F^\mu$ is the dual tensor, $^*F^\mu=1/2
\epsilon^{\mu\nu\rho}F_{\nu\rho}$.

In the Coulomb gauge we can define a field $\varphi$ such that
\be
A^i=\epsilon^{ij}\frac{\partial_j}{\sqrt{-\nabla^2}}\varphi ~~~.
\label{phi}
\ee
Note that the relation between the gauge field and  $\varphi$
is local in time, therefore the asymptotic theories for  $A_i$ and
$\varphi$ are equivalent.

Replacing  (\ref{phi}) in (\ref{dual}) we obtain the equation satisfied
by $\varphi$
\be
f(\Box)\varphi=0\makebox[.5in]{,}f(\Box)=C_1^{-1}(C_1^2(\Box)\Box+C_2^2(\Box))
{}~~,
\label{phinl}
\ee

The local Maxwell-Chern-Simons case would be associated with $C_1=1$,
$C_2=\mu$. In that case, eq.\ (\ref{phinl}) is the
usual Klein-Gordon equation, $f^{-1}(k^2)=1/(k^2-\mu^2)$ and
$\delta^+G=\delta(k^2-\mu^2)\theta(k_0)$. This represents an asymptotic state
with topological mass $\mu$ \cite{templeton}.

In the non-local case we are considering, replacing (\ref{1.20}) in
(\ref{phinl}), we obtain (taking $|u_+|=|u_-|=1$)
\be
f(\Box)=1/F(\Box)~~~.
\ee
($F(\Box)$ is defined by (\ref{1.10}), with $-k_E^2\rightarrow k^2$).
Therefore, the mass weight function for the modes present in the field results
\begin{eqnarray}
\lefteqn{\delta^+ G (k^2)= -i\, \left[
F(-(k_0+i\epsilon)^2+{\bf k}^2)-F(-(k_0-i\epsilon)^2+{\bf k}^2)\right] \theta
(k_0)}\nonumber \\
&&=-i\, \left[ F(-k^2-i\epsilon))-F(-k^2+i\epsilon)\right] \theta (k_0) ~~~.
\label{inv}
\end{eqnarray}

Now, we would like to make the following remarks:

Firstly, it is interesting that these expressions do not depend
on $G(\Box)$, so they do not depend on the (arbitrary) regularization
parameter $q$ used in the loop calculation; in particular, the mass weight
function is free from regularization ambiguities coming from the parity
non-conserving part of the polarization tensor.

Secondly, we have seen that the asymptotic states in the trivial sector can be
read from the $\delta$-singularities of the weight function (\ref{inv}), that
is, the poles of $F(k^2)$ (cf.\ (\ref{inv})). As $F(k^2)$ is the one-loop
parity conserving part of the polarization tensor, the presence of a pole in
this diagram could only come from a fermion-antifermion bound state in an
external electromagnetic field. Then, the only asymptotic state in the trivial
sector of the bosonized theory is the vacuum, as we know that there is no such
bound state, $F(k^2)$ in eq.\ (\ref{1.10}) has no poles.

Finally, the weight function then comes from the continuum of singularities
(cut) of $F(k^2)$, which leads to
\begin{equation}
\delta^+G=\frac{1}{8}\left( 1+\frac{4m^2}{k^2}\right) (k^2)^{-1/2}
\theta (k^2-4m^2) \theta (k_0)
\end{equation}
This is a positive definite function. This positiveness can be understood from
eq.\ (\ref{inv}): using unitarity, the mass weight function we obtained equals
a cross section for the production of a fermion-antifermion pair, which is of
course positive. Then, we see that the hamiltonian for this non-local theory,
unlike the higher order case, is bounded from below (cf.\ (\ref{Hrq})); so this
class of models are expected to display a sensible behavior. This physical
behavior can be traced back to the fermionic theory that originated the
non-local bosonized action.

Note that although the only asymptotic state in the trivial sector is the
vacuum, there would be, in general, positive norm states of the theory created
from the vacuum by non-polynomial functions of the fundamental fields, in an
analogous manner to the Mandelstam operators.

For example, in the $m=0$ case, equation (\ref{inv}) reduces
to
\be
f^{-1}(-k^2)=\frac{1}{16}\frac{1}{\sqrt{-k^2}}~~~,
\label{m0}
\ee
and the weight function reads
\be
\delta^+G=\frac{1}{8} (k^2)^{-1/2}\theta (k^2) \theta (k_0)~~~.
\ee
Here, the unique state in the trivial sector is the vacuum; however, this model
presents non-trivial (topological) fermionic excitations which are created from
the vacuum by non-polynomial operators, as described in Ref.\ \cite{marino}.

Summarizing, we have followed the Schwinger method, showing that it is possible
to quantize field theories having non-local kinetic terms in a way similar to
the local case.

A general non-local field operator presents a set of massive modes with weight
given by the discontinuity ($\delta^+G$) of the inverse kinetic operator. Each
mode contributes to the energy with weight $\delta^+G$.

The Fourier operators appearing in the mode expansion of the field are required
to obey commutation relations in such a way that the hamitonian be the
generator of time translations.

The Fourier operators $a^{\dagger}(k)$ are not, in general, creation operators
of (mass $k^2$) particles.
We have seen that the only massive modes associated with asymptotic massive
particles in the trivial sector are those related with $\delta$-singularities
in the weight function. In particular, we have analized a non-local
Maxwell-Chern-Simons model, showing that the only asymptotic state in the
trivial sector is the vacuum. This is related to the fact that fermions in
$2+1$ dimensions (interacting with an external field) have no bound states.

Finally, we would like to note that if we start from a physically well defined
local theory and generate a model containing non-local kinetic terms, the good
behavior of the initial theory should be manifested in the positiveness of the
weight function in the final model.
This occurs in effective models (see Ref.\ \cite{psdif}) and in the models
coming from bosonization in higher dimensions we have discussed in this letter.
In this last case, the stability (energy bounded from below) of the bosonic
theory is related to the positive metric of the Hilbert space of states in the
associated fermionic theory.

\section*{Acknowledgements}
This work was partially supported by Centro Latinoamericano de F\'\i sica
(CLAF), Conselho Nacional de Desenvolvimento Cient\'\i fico e Tecnol\'ogico
(CNPq), Brasil.

\end{document}